\documentclass[12pt]{article}
\usepackage{geometry}             
\usepackage{graphicx}
\usepackage{amssymb}
\usepackage{amsmath}
\usepackage{epstopdf}

\usepackage[hyperindex=true,
          pdfstartview=FitH,
          bookmarksnumbered=true,
          bookmarksopen=true,
          citecolor=blue,
          linkcolor=blue,
          colorlinks=true,
          unicode]{hyperref}

\begin{document}

\title{Accelerating BTZ spacetime}
\author{Wei Xu$^{1}$, Kun Meng$^{1}$ and Liu Zhao$^{1,2}$\thanks{email: {\it lzhao@nankai.edu.cn}}\\
${}^{1}$School of Physics, Nankai university, Tianjin 300071, China\\
${}^{2}$Kavli Institute for Theoretical Physics China, \\
CAS, Beijing 100190, China
}
\date{\today}                             
\maketitle

\begin{abstract}
An exact solution of $(2+1)$-dimensional Einstein gravity with 
cosmological constant is studied. The corresponding spacetime is interpreted as 
an accelerating BTZ spacetime. The proper acceleration, horizon structure, temperature
and entropy are presented in detail. The metric being studied 
is very similar to the one studied by Astorino in arXiv:\href{http://www.arXiv.org/abs/1101.2616}{{\tt 1101.2616}}, but the range of parameters is different which results in significant changes in the causal structures.
\vspace{4mm}

\noindent {\bf Keywords:} BTZ spacetime, proper acceleration, causal structure

\noindent {\bf PACS:} 04.20.Gz, 04.20.Jb
\end{abstract}

\section{Introduction}

In $n$-dimensional Einstein gravity, the static, spherically 
symmetric, vacuum black hole solution  takes the form
\begin{align}
\mathrm{d}s^{2}=
-\left(1-\frac{m}{r^{n-3}}-\epsilon\,\frac{r^{2}}{\ell^{2}}\right)\mathrm{d}t^{2}
+\left(1-\frac{m}{r^{n-3}}-\epsilon\,\frac{r^{2}}{\ell^{2}}\right)^{-1}
\mathrm{d}r^{2}
+r^{2}\mathrm{d}\Omega_{n-2}^{2}, \label{BH}
\end{align}
where $m$ is a mass parameter (not necessarily equal to the mass), 
$\epsilon$ is the sign of the cosmological constant 
\begin{align*}
\Lambda = \frac{(n-1)(n-2)\epsilon}{2\ell^{2}}
\end{align*}
and $\mathrm{d}\Omega_{n-2}$ is the line element of an $(n-2)$-sphere. 
It is known that generalizations of the above metric to include the cases with 
non-spherical, but maximally symmetric sections exist \cite{Mann, Birmingham:1998p1113}, 
i.e. the so-called topological black holes with metrics
\begin{align}
\mathrm{d}s^{2}=
-\left(k-\frac{m}{r^{n-3}}-\epsilon\frac{r^{2}}{\ell^{2}}\right)\mathrm{d}t^{2}
+\left(k-\frac{m}{r^{n-3}}-\epsilon\frac{r^{2}}{\ell^{2}}\right)^{-1}
\mathrm{d}r^{2}
+r^{2}\mathrm{d}\Sigma_{n-2,k}^{2}, \label{topoBH}
\end{align}
where $k=0,\pm 1$, $\mathrm{d}\Sigma_{n-2,k}^{2}$ is the line element of 
an $(n-2)$-dimensional maximally symmetric manifold which can 
be written explicitly as 
\begin{align}
\mathrm{d}\Sigma_{n-2,k}^{2}&=\mathrm{d}\theta_{1}^{2}
+ \rho_{k}^{2}(\theta_{1}) \mathrm{d}\Omega_{n-3}^{2},
\qquad \rho_{k}(\theta_{1})=\left\{
\begin{array}{ll}
\sin\theta_{1} & (k=+1) \cr
\theta_{1} & (k=0)\cr
\sinh\theta_{1} & (k=-1)
\end{array}
\right.  \label{rho}
\end{align}
where $\mathrm{d}\Omega_{n-3}^{2}$ is the line element of an $(n-3)$-sphere. 
The choices $k=0, -1$ are allowed only for $\epsilon=-1$, i.e. the asymptotically AdS
case. For all 
dimensions $n>3$, the metrics (\ref{topoBH}) are distinct for different values of $k$. 
However, for $n=3$, i.e. the case of BTZ black holes, there is some ambiguity, 
because there is no difference\footnote{Though, according to 
(\ref{rho}), the range of the coordinate $\theta\equiv\theta_{1}$ is different 
for different choices of $k$ in $n>3$ dimensions, one can always make it to stay in 
the range $[-\pi, \pi]$ in $n=3$, since in this case the metric becomes translationally
invariant along $\theta$, and we can identify points by taking the quotient 
of the group of this continuous translation symmetry by one of its discrete 
subgroup.} between $\mathrm{d}\Omega_{1}^{2}$ and 
$\mathrm{d}\Sigma_{1,k}^{2}$ and one can absorb $k$ 
into the mass parameter $m$ (as is usually done in the literature, see 
\cite{Banados:1992p1119} for instance).   

The study of Einstein gravity in three spacetime dimensions with cosmological constant
was initiated in \cite{Jackiw}. In the absence of matter source, this theory is locally 
trivial because of the lack of 
propagating degrees of freedom. In spite of this fact, there can be nontrivial boundary 
degrees of freedom and these are very interesting 
thanks to the AdS/CFT duality: the CFT dual of 3D 
gravity is $(1+1)$-dimensional, with infinite conformal symmetries making it completely 
integrable \cite{Brown:1986p547}. 
In the light that higher dimensional CFTs are often mathematically 
difficult to treat, one often take three dimensional gravity as an ideal toy model
for learning essential properties of the AdS/CFT duality. In this regard, various 
generalizations of BTZ black hole spacetime and their holographic properties have been 
studied extensively (see \cite{Anninos:2008p1297,
Clement:2009p4037,Chemissany:2009p3960,Chen:2010p5634,Tonni:2010p5733,
Chen:2011p1285,Song:2011p976} for a very incomplete list of related works).

In a recent paper \cite{Astorino:2011p1101}, 
Astorino studied an accelerating variant of the BTZ black hole. 
Basically, the spacetime metric obtained in \cite{Astorino:2011p1101} 
is a conformally transformed version of the 3D version of the metric 
(\ref{BH}), though the notations used are slightly different. This accelerating BTZ 
spacetime enjoys some feature which are otherwise absent in the non-accelerating 
version, e.g. the presence of black holes with an angular singularity, allowing 
extensions to black string of finite length and black ring in 4-dimensions, etc. 
Some properties of the spacetime 
are analyzed in detail in \cite{Astorino:2011p1101}, including horizon structure, 
temperature and entropy of the black holes in certain range of the parameters. 
In this paper, we shall extend the work of \cite{Astorino:2011p1101} and analyze 
the accelerating BTZ spacetime, allowing the acceleration parameter
to take a different sign from that in \cite{Astorino:2011p1101}. 
We also prefer to rewrite the metric of in a form which is conformal to the 
3D version of (\ref{topoBH}), taking advantage of the ambiguity in choosing 
parameters in the 3D case mentioned above. 

\section{Metric and proper acceleration}

We begin by writing  the spacetime metric in the following form,
\begin{align}
&  \mathrm{d} s^2=\frac{1}{\gamma(r,\theta)^{2}}
  \left[-f(r) \mathrm{d} t^2+f(r)^{-1}\mathrm{d}r^{2}
  +r^2 \mathrm{d} \theta^2 \right], \label{sol}
\end{align}
where 
\begin{align}
&  \gamma(r,\theta)=\alpha\, r\cosh(\sqrt{m-k}\,\theta)-1, \label{gamma}\\
&  f(r)=k-m-\epsilon\,\frac{r^2}{\ell^2},
\end{align}
where $k=0,\pm 1$, $\epsilon=0,\pm 1$ and $\ell$ is a positive 
parameter. Clearly, this 
is an exact solution to the 3D vacuum Einstein with cosmological constant 
$\Lambda$, which is given in terms of the parameters as 
\begin{align}
\Lambda=\alpha^{2}(m-k)+\frac{\epsilon}{\ell^{2}}. \label{cc}
\end{align}
For comparison, we also rewrite the original metric presented in 
\cite{Astorino:2011p1101} as follows:
\begin{align*}
\mathrm{d} s^2&=\frac{1}
{\left(1+\alpha\,r\,\cos\left(\sqrt{1-m}\,\theta\right)\right)^{2}}\bigg\{
-\left[1-m+r^{2}\left[\alpha^{2}(m-1)-\Lambda\right]\right]\mathrm{d} t^2\\
&\qquad +\left[1-m+r^{2}\left[\alpha^{2}(m-1)-\Lambda\right]
\right]^{-1}\mathrm{d}r^{2} + r^{2}\mathrm{d} \theta^2 \bigg\}.
\end{align*}
There are several differences in our presentation 
(\ref{sol}) of the metric from the one given in \cite{Astorino:2011p1101}.
First, we allow $k$ to take 3 possible values $0,\pm1$ rather 
than fix it to the value 1. The reason for this is to stress that for all 3 choices of values 
of $k$, the $m=0$ limit of the metric are just Einstein {\em vacua}, 
rather than black hole 
spacetimes \cite{Zhao:2011p693, XMZ} (for $k=\pm 1$, the corresponding vacua are
accelerating vacua, see below. Among these, the $k=1, m=0, \epsilon=1$ vacuum 
was first found and analyzed in \cite{Abn}.). Besides this purpose, there is no need to 
distinguish between $k$ and $m$ and the combination $m-k$ can be viewed as a single 
parameter. Second, we made a change $\cos(\sqrt{k-m}\,\theta)\rightarrow
\cosh(\sqrt{m-k}\,\theta)$. Though this is an identity transform, the use of hyperbolic
cosine reflects the actual angular behavior of the conformal factor in the 
presence of black hole which needs $m>k$, see below. The third and perhaps the most
significant difference lies in that the parameter $\alpha$ in our notation 
is actually $-\alpha$ in \cite{Astorino:2011p1101}. In the analysis of 
\cite{Astorino:2011p1101}, an implicit assumption of $\alpha>0$ was used (i.e. 
$\alpha<0$ in our notation), so we shall be dealing only with 
the choice $\alpha>0$. 

The coordinate ranges for $t$ and $r$ are specified as follows: $-\infty <t < \infty$, 
$0\le r <\infty$.  As for the coordinate $\theta$, due to the lack of translational 
symmetry in $\theta$ in the presence of the conformal factor, we seem to 
have no reason to restrict it to be in $[-\pi,\pi]$. However, we still wish to 
interpret the coordinates $(r,\theta)$ as being polar, just as in the standard case 
of BTZ black hole \cite{Banados:1992p1119}, since the latter is the $\alpha=0$
limit of our solution (\ref{sol}). Thus we make the choice $-\pi \le \theta\le \pi$.  
We shall see later that in some cases the range for $\theta$ still need to be adjusted 
for observers living in one side of the conformal infinity. 
The necessity of adjusting the range of $\theta$ in certain cases is one of the major 
point we would like to make with our choice of $\alpha$, in contrast to the cases 
analyzed in \cite{Astorino:2011p1101}.

Let us recall the explanation of the conformal factor $1/\gamma^{2}$ in the 
metric (\ref{sol}). Consider a static observer
\begin{align*}
x^\mu(\lambda) = \left(\lambda\frac{\alpha\, r\cosh(\sqrt{m-k}\,\theta)-1}
{\sqrt{k-m-\epsilon\,\frac{r^2}{\ell^2}}},r,\theta\right)
\end{align*}
in the spacetime, where $\lambda$ is the proper time.  It is easy to see that the proper 
acceleration $a^\mu =  u^\nu\nabla_\nu u^\mu$ (where $u^\mu=\frac{\mathrm{d}
x^\mu}{\mathrm{d}\lambda}$) has a norm equal to
\begin{align*}
a^{\mu}a_{\mu}=-\Lambda+\left(\frac{\epsilon}{\ell^2}\right)\frac{(k-m)
(\alpha \,r\cosh \left(\sqrt{m-k}\, \theta \right) -1)^{2}}
{k-m-\epsilon\,\frac{r^2}{\ell^2}}.
\end{align*}
At $r=0$, this reduces to
\begin{align}
a^{\mu}a_{\mu}=\alpha^{2}(k-m). \label{an}
\end{align}
This shows that $\alpha$ is {\it proportional (but not equal) to} the magnitude $|a|$ of the proper acceleration at the origin if $m\ne k$. Thus $\alpha$ may be called an acceleration 
parameter. Let us remark that the $t$-component of $a^{\mu}$ is zero. Because of this, 
$a^{\mu}$ is spacelike and we ought to have $a^{\mu}a_{\mu}>0$ in the 
static region of the spacetime. Combining with (\ref{an}), we see 
that $r=0$ is not in the static region if $m>k$. That the region of spacetime containing 
the origin $r=0$ is non-static is typical for black holes centered at the origin. 
For this reason, we infer that an accelerating black hole can exist only for $m>k$. 

\section{Horizon, temperature and entropy}

Now let us study the horizon structure. In general, zeros of $f(r)$ correspond to 
horizons. In our case the zeros are located at
\begin{align*}
r_{H}=\ell\sqrt{\frac{k-m}{\epsilon}}.
\end{align*}
This excludes, in particular, the possibility of choosing $\epsilon=0$. For a 
real zero to exist, we need either (i) $\epsilon=+1$ with $m<k$, or (ii) 
$\epsilon=-1$ with $m>k$. 

Notice that, according to (\ref{cc}), the sign of 
$\epsilon$ alone cannot determine whether the solution is asymptotically de Sitter
or anti-de Sitter.  We can re-express the cosmological constant in terms of $r_{H}$
as
\begin{align*}
\Lambda&=\frac{\epsilon}{\ell^{2}}(1-\alpha^{2}r_{H}^{2}).
\end{align*}
Therefore, for $\epsilon=+1$, we need $\alpha\, r_{H}>1$ to make an AdS background, 
$\alpha\, r_{H}<1$ to make a dS background, whereas for $\epsilon =-1$, 
$\alpha\, r_{H}>1$ makes a dS background and $\alpha\, r_{H}<1$ makes an AdS 
background. For either choices of $\epsilon$, $\alpha\, r_{H}=1$ corresponds to
a flat background.

The choice (i) (i.e $\epsilon=+1$, $m<k$) corresponds to 
the case in which static observers are located inside the horizon (i.e. $r<r_{H}$), 
hence the horizon is more like a cosmological horizon rather than a black hole 
horizon. Note that, in this case, the hyperbolic cosine function 
$\cosh(\sqrt{m-k}\,\theta)$ in the conformal factor is actually 
the usual  trigonometric cosine $\cos(\sqrt{k-m}\,\theta)$. Since 
$a^{\mu}a_{\mu}>0$ at $r=0$, the static region in this spacetime 
is surrounded by an accelerating horizon at $r=r_H$ which is not a black hole.

The choice (ii) ($\epsilon=-1$, $m>k$)
corresponds to the case in which static observers are located outside the horizon
(i.e. $r>r_{H}$), which is the case for a black hole solution. so we shall be 
concentrating only on this choice and bear in mind
\[
r_{H}=\ell\sqrt{m-k},\qquad \Lambda=\frac{1}{\ell^{2}}(\alpha^{2}r_{H}^{2}-1)
\]
from now on.

Notice that for the black hole case with $\epsilon=-1, m>k$, the 
cosmological constant $\Lambda$ needs not to be negative. This seems to be in
contradiction with the known no-go theorem proposed by Ida \cite{Ida}, which 
roughly says that the 
existence of a black hole horizon in $(2+1)$ dimensions requires a negative cosmological 
constant. The reason that our solution evades this no-go theorem is because that the 
proof of the theorem depends on the assumption that the curve corresponding to the 
horizon is everywhere smooth, however in our 
case the horizon contains an angular singularity along $\theta=\pm \pi$, which 
represents a string or strut acting on the horizon as the source of the 
acceleration. Similar situations have also been encountered in 
\cite{Astorino:2011p1101}. We will come back to this point later.

There are a number of standard ways to calculate the temperature at the horizon. For 
instance, we may take the timelike Killing vector $\chi=\partial_{t}$ to get the
surface gravity $\kappa = \sqrt{-\frac{1}{2}\nabla^{\mu}\chi^{\nu}
\nabla_{\mu}\chi_{\nu}}$ and then using $T=\frac{\kappa}{2\pi}$ to evaluate the 
temperature. The result reads
\begin{align*}
\kappa=\frac{\sqrt{m-k}}{\ell},\qquad T=\frac{\sqrt{m-k}}{2\pi\ell}.
\end{align*}
Another simple shortcut to the evaluation of the horizon temperature is to make use 
of the formula
\begin{align*}
T=\frac{1}{4\pi}\left.\sqrt{\frac{\mathrm{d} g_{00}}{\mathrm{d}r} 
\frac{\mathrm{d}(g_{11}^{-1})}{\mathrm{d}r}}\right|_{r=r_{H}},
\end{align*}
which is valid for any diagonal metric.

Now let us look at the function $\gamma(r,\theta)$ (eq.(\ref{gamma})) 
appearing in the conformal factor. Every zero of $\gamma(r,\theta)$
represents a conformal infinity in the metric (\ref{sol}). 
If the the acceleration parameter $\alpha$ were negative, then $\gamma(r,\theta)$
will never have a zero for $r>0$, provided 
$m>k$. So one does not need to worry about the conformal infinities. 
This is the case for \cite{Astorino:2011p1101}. In our case, however,  
conformal infinity does exist because we are interested in the choice $\alpha>0$. 
Moreover, the conformal infinity can intersect with the horizon if
$0<\alpha r_{H}\le 1$, i.e. in the AdS or flat cases.
This can be seen from the fact that the equation $\gamma(r_{H},\theta)=0$ has two 
solutions $\theta=\pm \theta_{0}$, with
\begin{align}
\theta_{0}=\frac{1}{\sqrt{m-k}}\,\mathrm{arccosh}
\left(\frac{1}{\alpha \,r_{H}}\right).  \label{t0}
\end{align}
If $\theta_{0}\in [0,\pi]$, then the intersection appears. From the outside static 
observer's point of view, this means 
that the horizon stretches all the way through infinity, i.e. it is noncompact. 
While evaluating the area of such horizons, one should exclude the part of 
the $r=r_{H}$ hyper surface that is hidden beyond the conformal infinity. 

Recall that the conformal 
infinity separates the spacetime into two patches: the $(-)$ patch 
$\gamma(r,\theta)\le 0$ and the $(+)$ patch $\gamma(r,\theta)\ge 0$. 
Each observer can perceive only one of the two patches. 
Therefore, the determination for the correct range for 
$\theta$ in the $\alpha\, r_{H}\le 1$ cases depends on which patch the observer 
lives in. For the AdS case with $\theta_{0}\in [0,\pi]$, we need to choose 
$\theta\in[-\theta_{0},\theta_{0}]$ in the $(-)$ patch and $\theta\in [-\pi, -
\theta_{0}) \cup (\theta_{0},\pi]$ in the $(+)$ patch. For the AdS case 
with $\theta_{0}>\pi$, 
the horizon will lie completely in the $(-)$ patch and we can still choose 
$\theta\in[-\pi,\pi]$. The flat case corresponds to $\theta_{0}=0$, the
horizon intersects with the conformal infinity at a single point $\theta=0$ in
the $(+)$ patch, in which case  we need to choose 
$\theta\in [-\pi, 0) \cup (0,\pi]$, and there is no horizon in the 
$(-)$ patch. For the de Sitter case, the 
zeros of  $\gamma(r,\theta)$ does not intersect with the horizon and we
still choose $\theta\in[-\pi,\pi]$. The presence of the conformal infinity makes the
analysis of the entropy and angular singularities of the horizon significantly different from the case of \cite{Astorino:2011p1101}, as is shown below.

First let us consider the entropy of the black hole spacetimes. We assume that  the Beckenstein-Hawking relation for the entropy still holds. 
For the asymptotically dS case (i.e. $\alpha\, r_{H}>1$).
The horizon lies entirely in the $(+)$ patch, and the entropy is
\begin{align*}
S=\frac{1}{4}\int^\pi_{-\pi}
\left.\sqrt{g_{\theta\theta}}\right|_{r=r_H}\mathrm{d} \theta=
\frac{\ell }{\sqrt{\alpha ^2 r_{H}^2-1}}
\mathrm{arctan}\left[\frac{\alpha  r_{H}+1}{\sqrt{\alpha^{2} r_{H}^{2}-1}}
\tanh\left(\frac{\pi}{2}\sqrt{m-k}\right)\right].
\end{align*}
For the asymptotically flat case, the horizon hits the conformal 
infinity at a single point $\theta=0$ in the $(+)$ patch, whereas there is no horizon in 
the $(-)$ patch. Therefore, we only need to evaluate the entropy in the $(+)$ patch, 
yielding a divergent result. 
The asymptotically AdS case must be 
subdivided into two subcases, i.e. $\theta_{0}\in [0,\pi]$ and $\theta_{0}>
\pi$. For $\theta_{0}\in [0,\pi]$, we have
\begin{align*}
\displaystyle
S&=\frac{1}{4}\int^{\theta_{0}}_{-\theta_{0}}
\left.\sqrt{g_{\theta\theta}}\right|_{r=r_H}\mathrm{d} \theta\\
&=\frac{\ell }{\sqrt{1-\alpha ^2 r_{H}^2}}
 \mathrm{arctanh}\left[\frac{1+\alpha  r_{H}}{\sqrt{1-\alpha^{2} r_{H}^{2}}}
 \tanh\left(\frac{1}{2} \mathrm{arccosh}
  \left(\frac{1}{\alpha r_{H}}\right)\right)\right]
\end{align*}
in the $(-)$ patch and
\begin{align*}
S&=\frac{1}{4}\left(\int^{-\theta_{0}}_{-\pi}+
\int^{\pi}_{\theta_{0}}\right)
\left.\sqrt{g_{\theta\theta}}\right|_{r=r_H}\mathrm{d} \theta\\
&=\frac{\ell }{\sqrt{1-\alpha ^2 r_{H}^2}}\left\{
 \mathrm{arctanh}\left[\frac{1+\alpha  r_{H}}{\sqrt{1-\alpha^{2} r_{H}^{2}}}
 \tanh\left(\frac{\pi}{2}\sqrt{m-k}\right)\right]\right.\\
&\qquad\left.
- \mathrm{arctanh}\left[\frac{1+\alpha  r_{H}}{\sqrt{1-\alpha^{2} r_{H}^{2}}}
 \tanh\left(\frac{1}{2}\mathrm{arccosh}\left(\frac{1}{\alpha r_{H}}\right)\right)
 \right]
 \right\}
\end{align*}
in the $(+)$ patch. For $\theta_{0}>\pi$, we have
\begin{align*}
S=\frac{1}{4}\int^\pi_{-\pi}
\left.\sqrt{g_{\theta\theta}}\right|_{r=r_H}\mathrm{d} \theta=
\frac{\ell }{\sqrt{1-\alpha ^2 r_{H}^2}}
\mathrm{arctanh}\left[\frac{1+\alpha  r_{H}}{\sqrt{1-\alpha^{2} r_{H}^{2}}}
\tanh\left(\frac{\pi}{2}\sqrt{m-k}\right)\right]
\end{align*}
in the $(-)$ patch and there is no horizon in the $(+)$ patch.

Next comes the analysis on the angular singularities of the horizon. Such singularities 
exists in all cases mentioned above because of the jump in the derivatives of $\gamma(r,\theta)$ at $\theta=\pm\pi$, and they provides the source of the acceleration 
which may be interpreted as a string or strut tied up to the horizon along the $\theta=\pm\pi$ direction. The tension $\tau$ of the 
string or strut may be calculated using the jump of the exterior curvature at the 
singularity using the formula \cite{Astorino:2011p1101}
\begin{align}
  [\mathcal {K}_{\mu\nu}]_{\theta=-\pi}^{\theta=\pi}-h_{\mu\nu}
  [\mathcal {K}]_{\theta=-\pi}^{\theta=\pi}=-h_{\mu\nu}\tau
  \label{jcond}
\end{align}
where $h_{\mu\nu}=g_{\mu\nu}-n_{\mu}n_{\nu}$ is the induced metric on the $
\theta=\pm\pi$ surface, $\mathcal {K}=\mathcal {K}_{\mu\nu}h^{\mu\nu}$, and $
\mathcal {K}_{\mu\nu}=\frac{1}{2}\mathcal {L}_{n}h_{\mu\nu}$ is the exterior 
curvature (with $\mathcal {L}_{n}$ representing the Lie derivative with respect to the 
normal vector $n^{\mu}$). Applying (\ref{jcond}) to the present case, we find that 
$\tau$ is negative in all cases except in the asymptotically AdS case with $\theta_{0}>
\pi$. Actually, the horizon surface stretches to the direction $\theta=\pm \pi$ in the $
(+)$ patch in all cases except the exceptional case mentioned above in which it goes to 
the $(-)$ patch. The explicit value of $\tau$ is
\begin{align}
\tau=-2\alpha\sqrt{m-k}\sinh(\pi\sqrt{m-k})
\end{align}
if the horizon reaches $\theta=\pm\pi$ in the $(+)$ patch, or
\begin{align}
\tau=2\alpha\sqrt{m-k}\sinh(\pi\sqrt{m-k})
\end{align}
if the horizon reaches $\theta=\pm\pi$ in the $(-)$ patch. The tension depends on the 
difference $m-k$ and it vanishes at $m=k$ or $m=0, k=1$. Besides the vanishing points, 
$\tau>0$ corresponds to a pushing strut and $\tau<0$ corresponds to a pulling string. 

Let us stress that, while doing the above analysis, we have not considered the possibility of extending the spacetime to the 
$r<0$ regime. If the latter regime were also included, then we will have an extra horizon
with $r_{H}<0$ as well as an extra conformal infinity at $r<0$ (if $\alpha$ were chosen 
negative). In any case the positive conformal infinity at $r>0$ and the negative 
conformal infinity at $r<0$ cannot be both present for a fixed $\alpha$. Note that this
last statement applies only to the black hole case ($m>k$). If we were considering the 
bubble spacetime with $m<k$, then the two conformal infinities will be both 
present for any given $\alpha$.

The mass or energy of the black hole solution is difficult to determine, because there is no known way to determine the mass of  asymptotically non-flat, accelerating black 
holes.  Nevertheless, since at $m=0$, the metric (\ref{sol}) becomes the accelerating 
vacua described recently by us \cite{XMZ}, 
we see that the mass of the black hole has to be 
proportional to $m$ by dimensional analysis\footnote{Remember that $m$ can actually
be written as $2GM$ where $G$ is the Newton constant and $M$ has the dimension 
of mass. Here we use the dimensionless mass parameter $m$ to make the notation 
simpler.}. One may, of course, assume that the first 
law of black hole thermodynamics holds in this situation and use that to calculate the 
mass of the black hole, as did in \cite{Astorino:2011p1101}. However, we 
prefer not to proceed in that direction due to the following reasons. One the one hand, 
the usual logic on the black hole thermodynamics is that the correctness of the first 
law of black hole physics ought to be checked after each 
observable of the black hole spacetime has been evaluated by other means, if 
gravity and black hole thermodynamic were considered as independent of each other. 
On the other hand, for spacetimes with non-vanishing cosmological constant (which is
the case of our main interests), the presentation of the first law contains some ambiguity 
even one is sure about its correctness. One may, for instance, consider the cosmological
constant as being the pressure of some liquid component and 
then write the first law in a form like $dE=TdS-pdV$ (see, e.g. 
\cite{Kastor:2009p4155}), rather than simply $dE=TdS$. In view of this, the first law 
alone is still insufficient to determine the mass  of the black hole, even if we accept the 
correctness of the first law by \textit{ad hoc} assumption.

\section{Extrinsic geometric description of the solution}

Unlike black hole solutions in dimensions $n>3$, the 3D black holes enjoys a very particular property, i.e. the Riemann curvature can be written as
\begin{align*}
R_{\mu\nu\rho\sigma}=\Lambda(g_{\mu\rho}g_{\nu\sigma}
-g_{\mu\sigma}g_{\nu\rho}).
\end{align*}
This allows us to embed some coordinate patch of 
the black hole spacetime into appropriate 4D flat spacetime
and study it from an extrinsic geometric point of view.

As in the previous section, we shall be working only with the choice (ii), i.e. 
$\epsilon=-1$, $m>k$, with the presence of black hole horizon in our solution.
We shall present the embedding relations according to different asymptotics of the
solution.

First we consider the asymptotically de Sitter case ($\alpha r_{H}>1$).
In this case, the spacetime can be embedded in a 4D Minkowski spacetime
\begin{align*}
  \mathrm{d} s^2=-(\mathrm{d} X_0)^2+(\mathrm{d} X_1)^2
  +(\mathrm{d} X_2)^2+(\mathrm{d} X_3)^2
\end{align*}
as a hyperboloid, 
\begin{align*}
  -(X_0)^2+(X_1)^2+(X_2)^2+(X_3)^2=\frac{\ell^2}{\lambda^{2}},
\end{align*}
where
\begin{align}  
 \lambda= \sqrt{\alpha^{2}\,r_{H}^{2}-1}=\sqrt{\alpha^2(m-k)\ell^2-1}.
\end{align}
The coordinate transformation leading to our solution (\ref{sol}) is given as
\begin{align}
  X_0&=\frac{1}{\gamma \sqrt{m-k}}
  \sqrt{r^2-(m-k)\,\ell^2}\,\sinh\frac{\sqrt{m-k}\,t}{\ell}, \label{emb1}\\
  X_1&=\frac{1}{\gamma \sqrt{m-k}}
  \sqrt{r^2-(m-k)\,\ell^2}\,\cosh\frac{\sqrt{m-k}\,t}{\ell},\\
  X_2&=\frac{1}{\gamma \sqrt{m-k}}
  \,r\sinh\left(\sqrt{m-k}\,\theta\right),\\
  X_3&=\frac{1}{\gamma \sqrt{m-k}}\,\lambda^{-1}
  \left(-r\cosh(\sqrt{m-k}\,\theta)
  +\alpha\sqrt{m-k}\,\ell^2 \right), \label{emb4}
\end{align}
where $\gamma=\gamma(r,\theta)$ is given in (\ref{gamma}). Notice that at $m=k$, 
the embedding breaks down, although the original metric (\ref{sol}) is well defined 
in that case, which corresponds to an accelerating AdS vacuum. Meanwhile, the 
embedding equations hold only for $r\ge r_{H}$, i.e. the static region of outside the 
horizon and the horizon itself, thus do not cover the whole spacetime.

The case of AdS black hole spacetime ($\alpha\, r_{H}<1$) can be embedded in 
a 4D pseudo Minkowski spacetime
\begin{align*}
  \mathrm{d} s^2=-(\mathrm{d} X_0)^2+(\mathrm{d} X_1)^2
  +(\mathrm{d} X_2)^2-(\mathrm{d} X_3)^2
\end{align*}
as a hyperboloid, 
\begin{align*}
  -(X_0)^2+(X_1)^2+(X_2)^2-(X_3)^2=-\frac{\ell^2}{\lambda^{2}}.
\end{align*}
The embedding relations are the same as (\ref{emb1})-(\ref{emb4}), but with 
$\lambda$ replaced by
\begin{align}  
 \lambda= \sqrt{1-\alpha^{2}\,r_{H}^{2}}=\sqrt{1-\alpha^2(m-k)\ell^2}.
\end{align}  

The flat case, $\alpha\, r_{H}=1$, is even more simpler: it can be seen that the 
asymptotically flat black hole metric
\begin{align*}
  \mathrm{d} s^2=\frac{1}{\gamma^2}\left[-(m-k)(-1+\alpha^2r^2) \mathrm{d} t^2
  +\frac{\mathrm{d} r^2}{(m-k)(-1+\alpha^2r^2)}+r^2 \mathrm{d} \theta^2 \right]
\end{align*} 
can be obtained from the 3D Minkowski metric
\begin{align*}
  \mathrm{d} s^2=-\mathrm{d}X_{0}^{2}+\mathrm{d}X_{1}^{2}
  +\mathrm{d}X_{2}^{2},
\end{align*}
thanks to the coordinate transformation
\begin{align*}
X_{0}&=\frac{1}{\gamma\alpha\sqrt{m-k}}\sqrt{\alpha^2r^2-1}
\sinh(\alpha(m-k)t),\\
X_{1}&=\frac{1}{\gamma\alpha\sqrt{m-k}}\sqrt{\alpha^2r^2-1}
\cosh(\alpha(m-k)t),\\
X_{2}&=\frac{1}{\gamma\sqrt{m-k}}r\sinh(\sqrt{m-k}\theta).
\end{align*}  
Let us stress once again that this extrinsic geometric description covers only a small 
portion of the black hole spacetime ($\alpha r >1$). It helps us to understand the flat 
nature of the spacetime, but it doesn't provides us with a global view of the spacetime.

The accelerating bubble spacetime corresponding to $\epsilon=+1$ and $m<k$ in
(\ref{sol}) can also be embedded into appropriate 4D target spacetimes 
following a similar spirit. 

\section{Causal structure of the spacetime}

The extrinsic geometric description of the metric can only describe a small coordinate patch of the spacetime which lies outside the horizon. To fully understand the 
structure of the spacetime, we identify its causal structure. As in the previous section, 
we shall be working only with the choice $\epsilon=-1, m>k$, i.e. black hole cases.

As usual, we introduce the Eddington-Finkelstein coordinates,
\begin{align}
  u=t-r^*, \quad v=t+r^*,
\label{E-F}
\end{align}
where the tortoise coordinate $r^*$ is defined as
\begin{align}
  r^*=\int (k-m+r^2/\ell^2)^{-1} \mathrm{d} r
  =\frac{\ell}{2\sqrt{m-k}}\log  \left|\frac{r-r_H}{r+r_H}\right|,\label{tortoise}
\end{align}
and both $u$ and $v$ belong to the range $(-\infty, \infty) $. In this coordinate the metric becomes
\begin{align}
  \mathrm{d} s^2 =\rho^2\left[-\left(k-m+\frac{r^2}{\ell^2}\right) \mathrm{d} u \mathrm{d} v 
  + r^2\mathrm{d}\theta^{2} \right],
\label{metric3}
\end{align}
where $\rho=\left(\alpha r\cosh(\sqrt{m-k}\,\theta)-1\right)^{-1}=\gamma(r,\theta)^{-1}$. 
The Kruskal-like coordinates are introduced as
\begin{align*}
\tilde{u}=\pm\exp\left(-\frac{\sqrt{m-k}}{\ell}u\right),\quad\tilde{v}=\pm\exp\left(\frac{\sqrt{m-k}}{\ell}v\right),
\end{align*}
where $\tilde{u}$ and $\tilde{v}$ takes the same sign if $r<r_H$,
and they take opposite signs if $r\geq r_H$.
So there are totally 4 different combinations,
each of which corresponds to a causal patch in the conformal
diagrams to be drawn below. In each cases, one finds that
\begin{align}
  \tilde{u}\tilde{v}=-\frac{r-r_H}{r+r_H},
\end{align}
and eq.(\ref{metric3}) becomes
\begin{align}
    \mathrm{d} s^2 = \rho^2
    \left[-\frac{(r+r_H)^2}{m-k}\mathrm{d} \tilde{u} \mathrm{d}
    \tilde{v} + r^2\mathrm{d}\theta^{2} \right],
\end{align}
where $r$ and $\rho$ are to be regarded as functions of
$\tilde{u}$ and $\tilde{v}$,
\begin{align*}
  r&=r_H\frac{1-\tilde{u}\tilde{v}}{1+\tilde{u}\tilde{v}},\\
  \rho&=\frac{1+\tilde{u}\tilde{v}}{\alpha r_H(1-\tilde u\tilde v)\cosh(\sqrt{m-k}\theta)-(1+\tilde{u}\tilde{v})}.
\end{align*}
Finally, the Carter-Penrose coordinates can be introduced by the
usual arctangent mappings of $\tilde{u}$ and $\tilde{v}$
\begin{align*}
  U&=\arctan{\tilde{u}},\quad  V=\arctan{\tilde{v}},\\
  T&\equiv U+V, \quad  R\equiv U-V,
\end{align*}
in terms of which the metric becomes
\begin{align}
 \mathrm{d} s^2 = \frac{\rho^2\ell^2}{\cos{R}^2}\left[- \mathrm{d} T^2 + \mathrm{d} R^2 
 +\frac{r^2}{\ell^2}\cos^2 (R) ~\mathrm{d} \theta^{2} \right].
\end{align}

The values of the product
$\tilde{u}\tilde{v}$ at $r=r_H$, $\rho=0$, $r=0$ and $\rho=\infty$ are respectively
\begin{align*}
 \lim_{r \rightarrow r_H} \tilde{u}\tilde{v}&=0,\quad 
 \Rightarrow UV=0\\
 \lim_{\rho \rightarrow 0} \tilde{u}\tilde{v}&=-1,\quad 
 \Rightarrow U-V=\pm\frac{\pi}{2}\\
 \lim_{r \rightarrow 0} \tilde{u}\tilde{v}&=1,\quad \Rightarrow 
 U+V=\pm\frac{\pi}{2}\\
 \lim_{\rho \rightarrow  \infty} \ \tilde{u} \tilde{v}&=
 \frac{\alpha r_H\cosh(\sqrt{m-k}\theta)-1}{\alpha r_H\cosh(\sqrt{m-k}\theta)+1}.
\end{align*}
These correspond to the various boundaries in the Carter-Penrose
diagrams. Among these, the first three sets of boundaries corresponding to 
$r=r_H$, $\rho=0$ and $r=0$ can easily be depicted, and in fact they constitute the 
Penrose diagram for the static, non-rotating and non-accelerating BTZ black hole 
\cite{Ortiz:2011p1293}. However, the forth set of the boundary, i.e. that 
corresponding to the conformal infinity $\rho=\infty$, depends on both the 
multiplication $\alpha r_{H}$ (i.e. on the cosmological constant) and the 
angular variable $\theta$. So we need to consider the causal diagrams separately for 
different choices of cosmological constant and the angle $\theta$.

\subsection{$\Lambda>0$}

\begin{figure}[h]
\begin{center}
\includegraphics[width=0.35\textwidth]{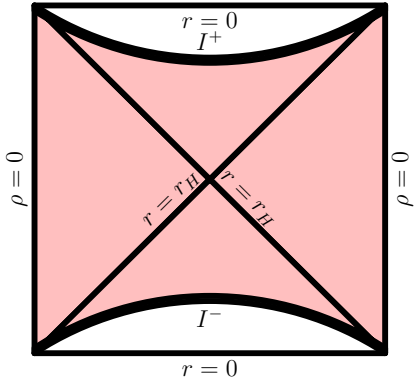}
\begin{minipage}{10.5cm}
\caption{Carter-Penrose diagrams for $\Lambda>0$: shaded area corresponds to the ($+$) patch, unshaded areas correspond to the ($-$) parch.} \label{Fig1}
\end{minipage}
\end{center}
\end{figure}

Since for $\Lambda>0$ we have $\alpha r_H\cosh(\sqrt{m-k}\theta)>1$,  
so $0<\lim_{\rho \rightarrow
 \infty} \tilde{u} \tilde{v}<1$. The Carter-Penrose diagram in this case
consists of the coordinate poles represented by the lines $\rho=0$, $r=0$, 
the horizon at $r=r_{H}$ and the (curved)  past and future conformal infinities at 
$\rho=\infty$. The corresponding diagram is depicted as in Fig.\ref{Fig1}.

In this and all subsequent figures, lines
labeled with $r=r_H$ represent acceleration horizons,
$I^+$ and $I^-$ are respectively past and future infinities ($\rho=+\infty$
or $r=\frac{1}{\alpha\cosh(\sqrt{m-k}\theta)}$), and $\rho=0$ or 
$I$ correspond to $r=+\infty$, the spacelike infinities.

\subsection{$\Lambda=0$}
\begin{figure}[h]
\begin{center}
\includegraphics[width=\textwidth]{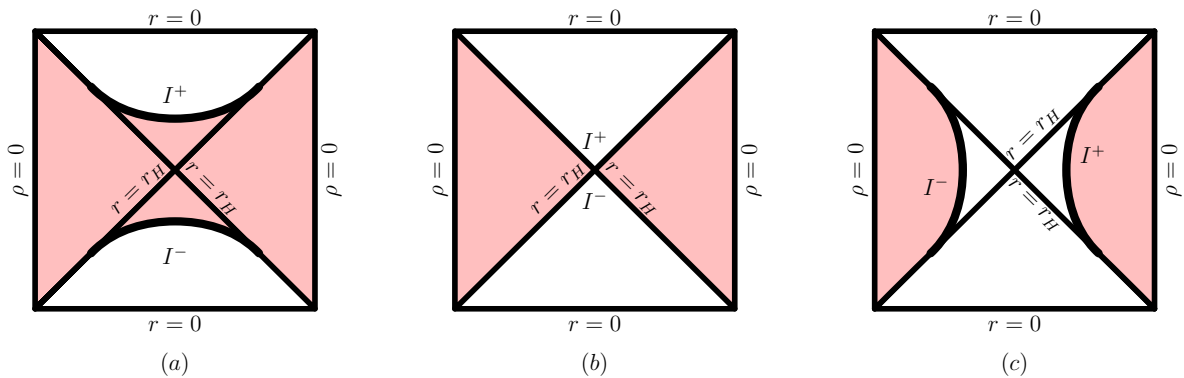}
\begin{minipage}{10.5cm}
\caption{Carter-Penrose diagrams for $\Lambda\leq 0$. } \label{Fig2}
\end{minipage}
\end{center}
\end{figure}

In this case, we have $r_H=\frac{1}{\alpha}\geq \frac{1}
{\alpha \cosh(\sqrt{m-k}\theta)}$, where equality holds only for 
$\theta=\theta_{0}=0$. The procedure for drawing Carter-Penrose diagrams 
for the case $\Lambda=0$ is very similar 
to the case $\Lambda>0$, except that we need to consider two subcases:

\begin{itemize}
\item $\theta\neq 0$ (i.e. $\theta\in[-\pi, 0)\cup(0, \pi]$): we have 
$0<\lim_{\rho\rightarrow\infty} \ \tilde{u} \tilde{v}<1$. The Carter-Penrose 
diagram is depicted in Fig.\ref{Fig2} (a);

\item $\theta=0$: in this case $\lim_{\rho\rightarrow\infty} \ \tilde{u} \tilde{v}=0.$ The Carter-Penrose diagram is depicted in Fig.\ref{Fig2} (b).
\end{itemize}

\subsection{$\Lambda<0$}

In this case, we have $r_H<\frac{1}{\alpha}$. The procedure to get the conformal 
diagrams is more involved because we need to consider two subcases
$\theta_0\in[0, \pi]$ and $\theta_0>\pi$.

\subsubsection{$\theta_0\in[0, \pi]$}
In this case,  
\begin{align*}
  \lim_{\rho \rightarrow\infty} \ \tilde{u} \tilde{v}&=\frac{\alpha r_H\cosh(\sqrt{m-k}\theta)-1}{\alpha r_H\cosh(\sqrt{m-k}\theta)+1}\\
  &=\frac{\cosh(\sqrt{m-k}\theta)-\cosh(\sqrt{m-k}\theta_0)}{\cosh(\sqrt{m-k}\theta)+\cosh(\sqrt{m-k}\theta_0)}.
\end{align*}
We need to subdivide the range of $\theta$ into three subcases:
$\theta\in[-\pi, -\theta_0)\cup(\theta_0, \pi]$, $\theta=\pm\theta_0$ and 
$\theta\in(-\theta_0, \theta_0)$.

\begin{itemize}
\item $\theta\in[-\pi, -\theta_0)\cup(\theta_0, \pi]$: 
we have again $0<\lim_{\rho\rightarrow\infty} \ \tilde{u} \tilde{v}<1$ and 
the Carter-Penrose diagram is similar to 
the $\Lambda>0$ case and is depicted in Fig.\ref{Fig2} (a).
\item  $\theta=\pm\theta_0$: in this case 
$\lim_{\rho\rightarrow\infty} \ \tilde{u} \tilde{v}=0$. 
The Carter-Penrose diagram is shown in Fig.\ref{Fig2} (b).
\item $\theta\in(-\theta_0, \theta_0)$: we have $-1<\lim_{\rho\rightarrow\infty} \ 
\tilde{u} \tilde{v}<0$. The Carter-Penrose diagram is depicted in Fig.\ref{Fig2} (c).
\end{itemize}

\subsubsection{$\theta_0>\pi$}
\begin{figure}[h]
\begin{center}
\includegraphics[width=0.35\textwidth]{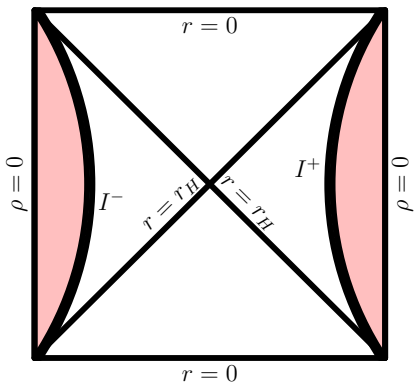}
\begin{minipage}{10.5cm}
\caption{Carter-Penrose diagrams for $\Lambda<0$.} 
\label{Fig3}
\end{minipage}
\end{center}
\end{figure}

In this case we have $-1<\lim_{\rho\rightarrow\infty} \ \tilde{u} \tilde{v}<0$.  
The Carter-Penrose diagram is depicted in Fig.\ref{Fig3}.

\section{Conclusion}

In this paper, we have made a detailed analysis of the accelerating BTZ spacetime
(\ref{sol}). The proper acceleration, horizon structure, temperature
and entropy are presented in detail. The presence of the accelerating parameter $\alpha$ 
has resulted in some new features comparing to the case of static BTZ spacetime. 
For instance, 
black hole horizon exists only in AdS background with $m>1$ in the static case, 
however,  in the presence of the acceleration parameter, black hole horizons exist for 
all possible signs of $\Lambda$, provided the parameter $\epsilon$ takes the value $-1$
and $m>k$ (we should take $m>0$ for $k=-1$ if positive mass is required). In addition, 
even if $\epsilon=+1$ we still have horizons which corresponds to accelerating bubbles
which are structures absent in the static case. 

Comparing to the case of 
\cite{Astorino:2011p1101}, our choice of the parameter range, especially the choice of 
sign for $\alpha$, has resulted in significant changes in the causal structure
of the spacetime. The unexpected rich structure in the causal diagrams makes the 
spacetime more interesting than the static BTZ black hole. 

In closing, let us mention that the accelerating BTZ spacetime discussed in this article
can be viewed as the 3D analogue of the C-metric in 4-dimensions. 
It is well known that the C-metric in 4D can be both charged and rotating
\cite{Pravda, Hong}. However, the 
solution we discussed in this article is only the analogue of the uncharged, non-rotating  
version. It will be interesting to ask whether one can find the accelerating version of 
charged and/or rotating BTZ spacetimes in 3 dimensions.
It is also tempting to ask whether 
the presence of proper acceleration can give rise to some novel features in the 
holographic dual picture. We leave these problems to future studies.

\section*{Acknowledgment} 

This work is supported in part by the National  Natural Science Foundation of 
China (NSFC) through grant No.10875059 and by the Project of Knowledge Innovation 
Program (PKIP) of Chinese Academy of Sciences, Grant No. KJCX2.YW.W10.


\providecommand{\href}[2]{#2}\begingroup\raggedright\endgroup

\end{document}